# A robust and reliable method for detecting signals of interest in multiexponential decays

Keith S Cover[1]
[1]Department of Physics and Medical Technology,
VU University Medical Center, Amsterdam

Corresponding author:
Keith S Cover, PhD                                     Email: Keith@kscover.ca
Department of Physics and Medical Technology           Tel:   31 20 444-0677
VU University Medical Center                           Tel:   31 20 444-0677
Postbus 7057
1007 MB Amsterdam
The Netherlands

**Running Title:** Robust multiexponential signal detection

## Abstract

The concept of rejecting the null hypothesis for definitively detecting a signal was extended to relaxation spectrum space for multiexponential reconstruction. The novel test was applied to the problem of detecting the myelin signal, which is believed to have a time constant below 40ms, in T2 decays from MRI's of the human brain. It was demonstrated that the test allowed the detection of a signal in a relaxation spectrum using only the information in the data, thus avoiding any potentially unreliable prior information. The test was implemented both explicitly and implicitly for simulated T2 measurements. For the explicit implementation, the null hypothesis was that a relaxation spectrum existed that had no signal below 40ms and that was consistent with the T2 decay. The confidence level by which the null hypothesis could be rejected gave the confidence level that there was signal below the 40ms time constant. The explicit implementation assessed the test's performance with and without prior information where the prior information was the nonnegative relaxation spectrum assumption. The test was also implemented implicitly with a data conserving multiexponential reconstruction algorithm that used left invertible matrices and that has been published previously. The implicit and explicit implementations demonstrated similar characteristics in detecting the myelin signal in both the simulated and experimental T2 decays, providing additional evidence to support the close link between the two tests. When the relaxation spectrum was assumed to be nonnegative, the novel test required signal to noise ratios (SNR) approaching 1000 in the T2 decays for detection of the myelin signal with high confidence. When the relaxation spectrum was not assumed to be nonnegative, the SNR requirements for a detection with high confidence increased by a factor of 25. The application of the test to a T2 decay from human white matter, measured in vivo with an SNR of 650, demonstrated a solid detection of the signal below 40ms believed to be due to the myelin water. The study demonstrated the robustness and reliability of extending the concept of rejecting the null hypothesis to relaxation spectrum space. The study also raised serious questions about the susceptibility to false positive detection of the myelin signal of the multiexponential reconstruction algorithms currently in use.

# I. INTRODUCTION

Multiexponential reconstruction has been used extensively in the reconstruction of relaxation spectra from T2 decays measured in vivo using MRI. However, as is well understood in the literature, a wide variety of relaxation spectra can be consistent with the same measured decay (1, 2, 3, 4, 5). This nonuniqueness is caused by both the noise in the measured decay and the finite number of points at which the decay was measured.

A central problem in detecting a signal arises when some of the relaxation spectra consistent with the data have the signal of interest and others do not. As the information in the data only provides sufficient information to assign a probability density, rather than a probability, to a relaxation spectrum (see the Theory section), it is impossible to assign probabilities to the existence of the signal of interest based on the information in the data alone. It is common practice in the literature to introduce information, in addition to that in the data, to assign probabilities (6 p 804). Such additional information is often referred to as prior information. In situations where spectra both with and without the signal of interest are consistent with the data, this additional information can determine whether the signal of interest is categorized as detected or not detected. If prior information is not available or unreliable, detection of the signal of interest becomes problematic. In particular, unreliable prior information can lead to a highly reproducible but false positive detection of signal, as will be demonstrated below.

To allow the detection of a signal of interest with or without the use of prior information, this paper introduces the novel statistical concept of extending the test of rejecting the null hypothesis to relaxation spectrum space. The null hypothesis in relaxation spectrum space is that the data is consistent with a relaxation spectrum that does not have the signal of interest. If the data can reject the null hypothesis to a high confidence level it follows that all the spectra consistent with the data have the signal of interest and the detection is definitive.

The detection of the myelin signal in in vivo T2 decays from the human brain will be used to demonstrate rejecting the null hypothesis in relaxation spectrum space both explicitly and implicitly. As is common practice in the MRI literature, a signal of interest below 40ms in the reconstructed relaxation spectrum will be considered to be from myelin water. Reliable measurement of the myelin signal, independent of the other water in the brain, would provide a means to study how the myelin is affected in diseases such as multiple sclerosis (7, 8) and schizophrenia (9). Therefore, we can reject the null hypothesis for the myelin signal, and thus provide a definitive detection, by showing a particular T2 decay is not consistent with zero signal below 40ms.

The fast Fourier transform (FFT), which is a form of the Fourier transform routinely used in spectral analysis, was the motivation for both the implicit and explicit form of the test of the test of rejecting the null hypothesis to relaxation spectrum space. The FFT can be represented as multiplication by an invertible matrix (or more generally a left invertible matrix (10)). The amplitude at a single frequency is considered definitely detected in a spectrum reconstructed by an FFT if it can be shown to be well above the noise. It is common practice to use the null hypothesis to test whether the amplitude is well above the noise (6 p 609). As the FFT has the properties of multiplication by an invertible matrix, the test of the null hypothesis can be transformed back from the frequency spectrum space to the original data (10). If the test is preformed on the original data it will be referred to as an explicit test. If it is performed on a single amplitude in a spectrum reconstructed from the original data it will be referred to as an implicit test.

Prior information that is often invoked in the reconstruction of multiexponential spectra from T2 decays is that the spectra are nonnegative. There is strong experimental and theoretical support for this assumption. And as will be demonstrated, this assumption greatly reduces the signal to noise ratio (SNR) required in a T2 decay to detect signals of interest.

Nonnegative least squares (NNLS) (11,12) is likely the most commonly used algorithm to reconstruct a relaxation spectrum from a multiexponential decay for myelin signal detection. NNLS finds the relaxation spectrum that has the least squares fit to the data with the added constraint that the relaxation is nonnegative. In practice, the NNLS algorithm without regularization usually produces relaxation spectra with a few very narrow peaks.

While the nonnegative assumption in NNLS has strong experimental and theoretical support, the least squares assumption does not. The least squares assumption, which is a special case of finding the relaxation spectrum with the smallest $\chi^2$ measure consistent with the data, has very little experimental support for multiexponential reconstruction (1). Thus, it is unclear how reliable reconstruction algorithms will be that combined both the nonnegative and least squares prior information.

The regularized form of NNLS it is widely used in the detection and measurements of the myelin signal (4, 13, 7). While regularization broadens the peaks slightly over nonregularized NNLS, regularized NNLS still tends to reconstruct spectra with the fewest number of peaks consistent with the data. Several publications have demonstrated its sensitivity to myelin signal detection (4, 5, 14) by showing its resistance to false negative detections. However, little has been published in the literature on its susceptibility to false positive myelin signal detection (10). This is unfortunate because methods that have maximum sensitivity tend to have less specificity, leaving such methods susceptible to false positive detections (15). While NNLS is the particular reconstruction algorithm used for comparison in this manuscript, other multiexponential reconstruction algorithms may also be susceptible to false positive signal detection. While this paper questions the reliability of some of the current practices in detection and measurement of the myelin water signal in T2 decays, it does not question the existence of the myelin water signal.

The rest of this paper compares the explicit and implicit forms of testing the null hypothesis by applying both tests to the myelin water detection problem for T2 decays, both simulated and measured in vivo, to demonstrate that both the implicit and explicit forms yield similar confidence levels of detection. The implicit forms used the data conserving multiexponential reconstruction matrices, which are left invertible, to reconstruct the relaxation spectra (10). The null hypothesis method is then compared to current practices in in vivo myelin water signal detection including the commonly used regularized NNLS algorithm.

A comparison of the data conserving multiexponential reconstruction algorithm and current practices including NNLS has been published previously (10). However, this comparison was for decays with a SNR 1000. For SNR of 1000, the worst problem found was that NNLS would reconstruct a low shoulder of the main water peak into a separate myelin peak. However, as this was a case of false peak detection rather than false signal detection, the problem was not particularly serious. The present paper considers the reconstruction problem were the SNR is only 100 which is closer to current practices in in vivo T2 measurements than 1000.

## II. THEORY

The multiexponential forward problem, as it is routinely stated, has the form

$$d_n = \int_0^\infty m^O(\tau) \exp(-t_n/\tau) d\tau + n_n \quad (1)$$

where $d_n$ is the data value sampled at time point $t_n$, n=1,...,N, and where N is the number of data points, $\tau$ is time constant, $m^O(\tau)$ is a continuous function called the original relaxation spectrum, and $n_n$ is the additive noise at time point $t_n$. The most common current practice in vivo MRI measurements is to measure the decay at 32 time points with a spacing of 10ms between each time point. The SNR of the T2 decays is usually defined as the value of the first point in the decay divided by the standard deviation of the noise. While reliable SNR estimates for T2 decays are difficult to find in the literature (16), from the author's experience and discussions with other researchers, under current practices in vivo measurement, the SNR is in the range of 100 to 300 (17).

The goal of the multiexponential reconstruction problem is to reconstruct $m^O(\tau)$ in Eq. 1 from $d_n$. As is well known in the literature, it is impossible to exactly reconstruct $m^O(\tau)$ given the limited information in the finite and noisy data points $d_n$. Still, even limited information about $m^O(\tau)$, such as whether we can state to a high confidence level that there is a signal of interest below 40ms, can be very useful. There is a wide variety of multiexponential reconstruction algorithms available in the literature. However, these algorithms can give quite different answers about the existence of a myelin signal when applied to the same multiexponential decay (1, 2, 3, 4).

The concept of ideal signal and noise will be used throughout this paper. The decay signal will be considered ideal when the original relaxation spectrum generating the signal is constant over time and free of artifacts. Noise will be considered ideal if it is Gaussian, stationary, uncorrelated, additive and has a mean of zero. The complications introduced into reconstruction due to non ideal signal, such as even echo rephasing, or non ideal noise, such as Riccian noise, is beyond the scope of this paper.

At the heart of most multiexponential reconstruction algorithms is a model spectrum's consistency with the data. The most commonly used measure of the misfit between measured data, $d_n$, and data yielded by a model spectrum, $d_n^M$, is the $\chi^2$ measure (6 p 806). The equation for calculating the $\chi^2$ is

$$(2)$$

where $\sigma_n$ is the standard deviation of the noise in measurement $d_n$ and the noise is assumed to be ideal.

If the proposed model spectrum is identical to the original spectrum, then the probability distribution for $\chi^2$ depends only on the statistics of the noise. Since the noise is assumed to be ideal, the probability distribution for $\chi^2$ is, to a good approximation for 32 or more data points, Gaussian. The distribution has a mean of N and a standard deviation of sqrt(2N), where N is the number of data points (6 p806). Thus, the mean and standard deviation of $\chi^2$ for 32 echoes are 32 and 8, respectively.

The above statistics for the $\chi$2 assumes the relaxation spectrum is a continuous function that has been approximated by a finite number of values for computational purposes. It should be appreciated that this is a very different assumption than assuming the finite number of parameters completely characterizes the relaxation spectrum. The latter assumption, which is referred to as the parameterized model assumption (6 p 689), does not accurately represent the multiexponential reconstruction problem and will yield incorrect statistics for the $\chi^2$. In particular, parameterized models modify the $\chi^2$ statistics by a "degrees of freedom" parameter. Since the relaxation spectrum is a continuous function, degrees of freedom should not be taken into account when calculating the $\chi^2$ statistics.

While it is possible to assign a probability distribution to $\chi^2$, provided the statistics of the noise are known, the information in the data alone does not provide sufficient information to assign a probability to each model relaxation spectrum. However, it does provide sufficient information to calculate a probability density from it's $\chi^2$ fit to the data (18),

$$\Pr obDen(\chi^2) = k \exp(-\chi^2 / 2) \qquad (3)$$

where $k$ is a normalization constant. The difference between probabilities and probability densities are often overlooked in discussions on reconstruction. You need additional prior information to turn a probability density into a probability. Unfortunately, the nonnegative prior information alone is generally insufficient information to determine probabilities. Thus, confidence levels can only be assigned for a specific $\chi^2$ and all model spectra with the same $\chi^2$ will have the same confidence level.

A useful property for characterizing the reconstruction matrices is the noise gain (10). If the noise of the measured data is ideal and has a known standard deviation it is easy to calculate the noise at each point in a reconstructed spectrum if the reconstruction algorithm is a matrix multiplication. For each row of a reconstruction matrix with coefficients $a_n$, the standard deviation of the noise of the corresponding point in the reconstructed spectrum is

$$\qquad (4)$$

A convenient value to define for each row of a reconstruction matrix is the noise gain, $G_N$. It is defined as

$$\qquad (5)$$

Provided all data points have ideal noise and equal standard deviations, the noise gain is the factor by which the noise is increased during the reconstruction by this particular row. One of the useful properties of the reconstruction matrices used in this paper is that all rows of a matrix have the same noise gain. This property yields reconstruction spectra where the standard deviation of the noise is constant over the whole spectrum.

## III. METHODS

The analysis in this study consisted of several steps. The first step was the generation of simulated T2 decays, for use as original spectra, to aid in the assessment of the various detection methods of the myelin signal. The second step demonstrated common practices in multiexponential reconstruction currently used in the detection of the myelin signal so they could be compared with testing the null hypothesis. In the second step, several versions of regularized NNLS were used to reconstruct simulated decays. The third and fourth steps evaluated the detection abilities of rejecting the null hypothesis both explicitly and implicitly. The explicit detection was conducted 1) with only the information in the data and 2) with the additional prior information that the relaxation spectrum was nonnegative. The implicit detection used the data conserving reconstruction multiexponential algorithm described in Cover (10) and assumed the nonnegative relaxation spectrum assumption only when interpreting the resulting relaxation spectrum.

As will be described below, a total of five versions of the NNLS algorithm were used in this paper. Three regularized versions, as described in the literature or slightly modified from the literature, were used to reconstruct T2 curves. A fourth version of NNLS, which was not regularized, was used to test the null hypothesis explicitly and will be referred to as the null hypothesis version. A fifth version of the NNLS, which was also not regularized, was used to design a original relaxation spectrum, referred to as the broad peak spectrum, that was susceptible to false positive myelin signal detection. The fifth version of NNLS will be referred to as the false negative version.

All optimization in this paper was performed using AMPL (20).

**A. Simulated original spectra**

Two original spectra were used to generate simulated T2 decays for this paper. The first was a simple model of a white matter relaxation spectrum found in the literature (17). In the original spectrum, the myelin and main water peaks were located at 20ms and 80ms respectively with the main peak having 9 times the amplitude of the myelin peak. This original spectrum will be referred to as the biexponential original spectrum.

The second original spectrum was intended to be a slightly more realistic model of the main peak than the monoexponential decay used in the biexponential original spectrum. It was also intended to be more difficult to reconstruct accurately. Thus the signal below 60ms was zeroed and the main peak was broadened in such a way that it might be mistaken for a myelin signal. The second original spectrum will be referred to as the broad peak spectrum.

The broad peak spectrum was generated using a version of NNLS that was inspired by the "false negative" version of NNLS in Cover (10). The version was trying to suppress the myelin signal that was known to exist in the simulated decay. The basic nonregularized NNLS was modified with the addition of several constraints. The additional constraints were intended to reconstruct a relaxation spectrum that was roughly similar to a main water peak.

The additional constraints were that 1) the relaxation spectrum below 60ms was constrained to zero, 2) the spectrum was constrained to increase monotonically from 60ms to 80ms and 3) decrease monotonically above 80ms. This version of NNLS was applied to 20 simulated T2 decays generated by the biexponential original spectrum and additive noise so the T2 decay has a SNR of 100. The only difference among the 20 decays was the noise. While maintaining the same standard deviation, the noise was different for each decay. The decays had 32 sample points with a spacing of 10ms. Of the 20 reconstructed spectra, the reconstructed spectrum with the least broad peak that was still a reasonable fit to the decay was selected for use as the broad peak spectrum. The least broad peak was selected because it best approximated a single main water peak.

Under current practices for measuring relaxation spectra in vivo, it is common to sample the T2 decays at 32 echoes with a 10ms spacing and then reconstruct using a regularized NNLS algorithm (14, 8, 13, 11). Therefore, for comparing of the data conserving reconstruction matrices to current practices, both the NNLS algorithm and the multiexponential reconstruction matrices will reconstruct decays 32 echoes at 10ms spacing.

**B. Reconstruction using NNLS**

Three versions of the regularized NNLS algorithm were then used to reconstruct the decays. The details of the 3 versions are in line with those commonly used in the literature. All three versions used the same regularization (11). First, the minimum $\chi^2$ was found by using NNLS without regularization. Then the trade-off parameter between the $\chi^2$ and relaxation spectrum parameter was adjusted so the $\chi^2$ value was increased by between 1% and 2% above the minimum.

The only difference between the 3 versions of regularized NNLS used in this paper was the range of time constants used in the relaxation spectrum. The first version used time constants ranging from 1 to 2000ms, the second version used time constants ranging from 15 to 2000ms, and the third version only used 4 discrete time constants at 20, 80, 120 and 2000ms. The first two versions approximate the continuous relaxation spectra with 20 evenly spaced monoexponential per decade. The latter two versions have been used in the literature in detecting and measuring the myelin signal with the second version being most commonly used (19, 11). These three versions are the same as used in Cover (10).

To assess the performance of the versions of the NNLS algorithm, 20 realization of the decay from the broad peak for SNR of 100 were generated. Twenty realizations were chosen because 20 was a large enough number to provide adequate statistics but small enough that when all 20 reconstructed relaxation spectra were plotted on the same graph they gave a reasonable indication of the range of the reconstructed spectra.

It is common in the literature to estimate the myelin signal from a relaxation spectrum by integrating a reconstructed relaxation spectrum over all signal below 40ms. For this paper, this value will be referred to as the "short time constant signal" (STCS).

**C. Testing the null hypothesis explicitly**

The null hypothesis was that there was no signal below 40ms in an original relaxation spectrum of which the T2 decay had been measured. The null hypothesis was explicitly tested in two different forms. The first form used only the information in the data, thus with no prior information. The second form included the additional prior information that the relaxation spectrum was nonnegative. Both forms of the test used standard reconstruction algorithms to find relaxation spectrum with the lowest $\chi^2$ fit to the decays. In both reconstruction algorithms, the relaxation spectrum was constrained to zero below 40ms.

The reconstruction algorithms differed in how they handled the relaxation spectra above 40ms. The form of the null hypothesis test with no prior information had no constraints on the spectra above 40ms. Thus the null hypothesis could be tested with a reconstruction algorithm that found the spectrum with the smallest $\chi^2$ fit to each decay. The null hypothesis test with the nonnegative spectrum constraint was implemented by the NNLS algorithm that, as was mentioned above, found the $\chi^2$ fit to each decay with the constraint the relaxation spectrum was nonnegative.

Both forms of the null hypothesis test were applied to 100 simulated T2 decays generated from the biexponential spectrum with a range of SNR. The decays had 32 echoes spaced at 10ms. For each SNR, the mean and standard deviation of the 100 $\chi^2$ were calculated. The statistics of the $\chi^2$ was then compared to the probability distribution function of the $\chi^2$ due to noise to determine the SNR required to reject the null hypothesis both with and without the nonnegative constraints.

**D. Testing the null hypothesis implicitly**

The null hypothesis was realized implicitly using the reconstruction matrices for the data conserving reconstruction algorithm as described in Cover (10). The decays were assumed to have been sampled at 32 evenly spaced echoes. In the calculation of the reconstruction matrices, the continuous relaxation spectra were approximated by monoexponential decays spaced at 50 time constants per decade ranging from 0.1ms to 10,000s. The reconstruction matrices were calculated for noise gains of 1, 3.16, 10, 31.6, 100, 316 and 1000. Both the point spread functions and resolution functions (10) for each of the reconstruction matrices were calculated and plotted.

The data conserving nature of the reconstruction matrices meant that no prior information was included in the reconstruction of the relaxation spectra. However, as will be demonstrated, the nonnegative constraint can be easily taken into account during the interpretation of the reconstructed spectra.

**E. *In vivo* T2 decay**

To demonstrate the null hypothesis test on a T2 decay measured in vivo, a T2 decay was measured from white matter in the internal posterior capsule of a human brain. The decay consisted of 32 echoes with a 10ms spacing followed by 16 echoes with a 100ms spacing. While not identical sampling to 32 echoes at 10ms, the sampling is sufficiently close to contain similar information about the myelin signal.

The *in vivo* T2 decay analyzed is the same one used in Cover (10) and a plot of the decay is presented in Fig. 7 of that paper. The SNR of the decay was estimated to be 650. As the decay measures the magnitude of the complex T2 signal, all values are positive. Thus, when the signal dwindles to values less than the noise, the measured values yield only positive values. Therefore, the measurements become nonlinear as the data values approach zero. As can be seen from the plot of the decay, the last 8 points are all positive and appear to represent primarily noise. To gauge the effects of the Raleigh distribution of the decay, both the first 32 echoes and the full 48 were tested for the existence of signal below 40ms using the null hypothesis explicitly. The implicit test was included in Cover (10) and plots of the relaxation spectrum that were calculated using data conserving reconstruction matrices can be found in the paper.

For this paper, the test of the null hypothesis was implemented by constraining the amplitude of all the time constants below a specified time constant to zero while finding the relaxation spectrum with the lowest $\chi^2$ value using the nonregularized version of NNLS. The specified time constants ranged from 1ms to 100ms. The relaxation spectrum was discretized at 20 points per decade from 1ms to 2s.

## IV. RESULTS

### A. The broad peak original spectra

Figure 1 shows the broad peak spectrum – one of two original spectra used to generate simulated decays for this paper. Key characteristics of the broad peak spectrum are that it has no signal below 60ms and only a very small signal above 90ms. The broad peak is wider than most of the original spectra used to assess the performance of multiexponential reconstruction algorithms for in vivo T2 decays (1, 4, 5, 14, 17).

The mean and standard deviation of the $\chi^2$ for the 20 reconstructed spectra from which the broad peak spectrum was chosen were 28±6. As these values fall within or below the range of $\chi^2$ values that are considered consistent with the data (32±8), there was little problem finding relaxation spectra consistent with the biexponential T2 decays that had no signal below 60ms.

### B. Reconstruction using NNLS

Figure 2 shows the NNLS reconstructions of simulated decays generated from the relaxation spectrum in Fig 1 with SNR of 100. The means and standard deviations of the $\chi^2$ from the 3 versions of regularized NNLS reconstruction were (a) 24±6, (b) 25±6 and (c) 28±7 respectively. NNLS will often find fits better than the expected mean for $\chi^2$ of 32 because of its nonlinear nature. As would be expected, the more restrictive the range of time constants in the relaxation spectrum the larger the $\chi^2$. However, all mean values are below the expected mean of $\chi^2$ indicating all 3 versions of the regularized NNLS had little problem finding relaxation spectra consistent with the data.

The STCS for the 3 versions of NNLS are (a) 0.042±0.049, (b) 0.014±0.019 and (c) 0.095±0.025 respectively. Thus the STCS also shows that the third version of the NNLS yields a highly reproducible myelin signal even though there was no myelin signal in the original spectrum that generated the T2 decay. The second version (b) shows effectively no myelin signal and the first version (a) shows a partial detection. While Fig. 2 gives the impression that there is signal below 40ms in (b) the amount of signal averaged over the 20 realizations is not significant. Thus, the three versions of the NNLS each yielded results that were highly reproducible but very different even though all three were consistent with the decays. Thus, the 3 different versions of NNLS yielded very different results for the STCS for the same decays. In particular, the spectra reconstructed with the third version of the regularized NNLS algorithm (c) shows highly reproducible myelin peaks even though there is none in the original relaxation spectrum (Fig 1.).

### C. Testing the null hypothesis explicitly

Table 1 gives the estimate of the confidence level at which the null hypothesis could be rejected for a range of SNR assuming no prior information was included in the myelin signal detection. The estimate of the confidence level was calculated by first subtracting off the expected $\chi^2$ and then dividing by the expected standard deviation. While the confidence levels should be positive, due to statistical fluctuations, occasionally they are negative. To be considered statistically significant, a null hypothesis has to be rejected by at least 2σ (15). The confidence level exceeded 2σ when the decay SNR exceeded 12,580.

Table 2 gives the confidence level at which the null hypothesis could be rejected for a range of SNR assuming the relaxation spectra was nonnegative. A T2 decay with an SNR of at least 500 was required to reject the null hypothesis at a confidence level of 2σ.

Thus, when decays are measured at 32 echoes spaced at 10ms and with the null hypothesis that the relaxation spectrum below 40ms is zero, the prior information that the relaxation spectrum is nonnegative reduce the SNR required to reject the null hypothesis by a factor of 25. Based on these simulations, and assuming a nonnegative relaxation spectrum, to detect significant changes in signal below 40ms, rather than just detecting its existence, will require SNR approaching 1000 or higher.

**D. The reconstruction matrices**

Figure 3 shows the point spread functions for each of the seven reconstruction matrices calculated. The point spread functions were generated by monoexponential decays with time constants of 10, 20, 50, 100, 200, 500 and 1000ms and plotted for each reconstruction matrix. The linearity of the reconstruction matrices combined with the shape of the point spread functions allow an interpreter to quickly make a rough estimate of how any original spectrum reconstructed by a particular reconstruction matrix will appear.

The point spread functions show an increase of resolution by more than a factor of 2 between the reconstruction matrix with the lowest and highest noise gains. However, noise gain ranges from 1 for the reconstruction matrix with the least resolution to 1000 for the reconstruction matrix with the highest resolution. Thus the doubling of the resolution comes at a very high cost in terms of SNR of the decays. This high cost has been previously reported in the literature (1, 10).

Figure 4 displays the resolution functions for the seven reconstruction matrices. Each resolution function was generated from its corresponding row in the reconstruction matrix and precisely characterizes its resolution. Each row in a reconstruction matrix also corresponds to a point in the reconstructed spectrum. Thus the resolution function gives the precise resolution of the corresponding point in the reconstructed spectrum.

Comparison of the resolution functions with the point spread functions of Fig. 3 shows a similar increase in resolution with noise gain. The roughly Gaussian shape of the resolution functions ensure that very little signal, other than that near the peak of the resolution function, contributes to its associated value in a reconstructed spectrum.

The resolution functions of the matrices presented in this paper have a few special properties that are worth keeping in mind. First, the area of the resolution functions for the reconstruction matrices presented in this paper are always unity. Thus, the value in a relaxation spectrum is a localized average of original spectrum that generated the decay. Second, the peak of the resolution function is always located at the time constant corresponding to the resolution function. This property ensures that the resolution function is averaging the correct range of time constants of the original spectrum. Thirdly, the resolution functions decrease monotonically from the peak and are always non zero. This property allows the resolution functions to perform a more effective job in resolving the original spectrum.

**E. Testing the null hypothesis implicitly using reconstruction matrices**

Figure 5 shows the spectra reconstructed with the data conserving reconstruction matrices from a decay generated by the biexponential original spectrum. The SNR of the decay was 1000. All of the spectra were reconstructed from the same decay. The difference between the reconstructions is that the noise gain ranges from 1.0 to 31.6. The standard deviation of the noise in each reconstructed spectrum is displayed as error bars to the left of each plot. The error bars show 1 standard deviation above the middle bar and one below. The standard deviation can be calculated reliably as we know the standard deviation of the ideal noise in the decay and the noise gain of each of the reconstruction matrices.

In Fig. 5 the main water peak stands out as it is much larger than the noise. However, care must be taken to reliably detect the smaller myelin signal. The reconstructed spectrum must have sufficient resolution so that the myelin signal does not substantially overlap with the signal from the main water peak. Examination of the point spread functions and resolution functions shows that the reconstructed spectra with G=10 and G=31.6 have sufficient resolution provided the original relaxation spectra are nonnegative. At 20ms both spectra have a positive signal. However, only for G=10 does the signal stand several standard deviations above the noise, clearly rejecting the null hypothesis in relaxation space, and yielding a definitive detection.

Figure 6 shows the reconstruction of the same biexponential decay as Fig. 5 but with SNR of the decays of 100 instead of 1000. Examination of the reconstructions for the various noise gains show a main peak standing well above the noise. However, any signal in the region of were the myelin signal is expected is clearly below the noise floor. In addition, the noise is higher than the amplitude of any expected myelin signal. Thus, at an SNR of 100, the data is too noisy to see the myelin signal.

For the simulated data used in this paper the SNR of the decays is known. However, in practice, estimating the SNR in vivo can be difficult. Reconstruction matrices provide a simple way to estimate the noise in a T2 decay provided ideal noise is assumed. The G=31.6 shows only noise below 30ms. It is possible to estimate the noise in the reconstructed spectrum from this interval and then divide by the noise gain to find the noise of the T2 decay. The noise estimate of the T2 decay can then be used to estimate the noise in another reconstructed spectra of the same T2 decay but with a lower noise gain.

Figure 7 shows the reconstructed spectrum for the broad peak original spectrum where the T2 decay had an SNR of 1000. It is clear from the G=10 reconstruction there is little or no myelin signal at 20ms. Figure 8 also shows a T2 decay from the same original spectrum however the SNR was 100. The reconstructions are much too noisy to say whether the T2 decay is consistent with a myelin signal or not.

**F. Analysis of the *in vivo* T2 decay**

Table 3 gives the values of $\chi^2$ for both the 32 and full 48 echo decays. From the theory section the expected mean and standard deviation for 32 and 48 point decays are 32±8 and 48±9.8 respectively.

When it was specified that all time constants below 1ms were to be forced to zero, the $\chi^2$ values are 28.8 and 192.0 respectively. While the $\chi^2$ value for 32 echoes is well within the expected range, the value for 48 echoes is too large. As mentioned in the methods section, this is likely due to the magnitude of the noise in the last echoes introducing a nonlinear component in to the data.

For 32 echoes, the $\chi^2$ value exceeds $2\sigma$ at 28.2ms and $4\sigma$ at 35.5ms. If, for the 48 echo decay, we assume the $\chi^2$ is offset by a constant factor of 144 (192-48), then it exceeds $2\sigma$ at 28.2ms and $4\sigma$ at 35.2ms. Thus the 32 echo and 48 echo yield similar rejections of the null hypothesis and both confirm the existence of a signal with a time constant below 40ms, which is believed to be due to myelin.

## V. DISCUSSION

### A. Reconstruction with NNLS

The standard deviation of the STCS indicates a highly reproducible myelin signal detection for the third version but no detection for the second version of the regularized NNLS. This is in spite of the fact that both versions reconstructed the same T2 decays. As mentioned above, both the second and third versions are used in the literature. The first version of the regularized NNLS yielded a partial detection. High reproducibility of signal in a reconstructed spectrum is often assumed to indicate a detection with high confidence. This assumption is true for the FFT and is also true for the spectra reconstructed with the multiexponential reconstruction matrices because of data conservation and the implicit rejection of the null hypothesis. However, the results of this paper clearly indicate that, for the NNLS algorithm, a highly reproducible signal in a spectrum does not ensure a reliable detection. NNLS detections can depend critically on the underlying least squares prior information. But if the prior information is unreliable the reconstructed spectrum can have a highly reproducible but false positive signal.

It could be asked how realistic is the broad peak original spectrum? A simple answer would be to say that it is more realistic than the monoexponential peak commonly used in the literature to model the main peak. For example, the main peak of the biexponential original spectrum used in this paper is a monoexponential. Another answer would be to point out this is only one example of an original spectrum that yields a false positive myelin signal. There could well be many more.

However, such arguments would ignore one of the main goals of this paper: when trying to detect a signal, to avoid prior information as much as possible when the information may be unreliable. The NNLS algorithm uses a pair of assumptions in addition to the regularization. It assumes the spectrum is both nonnegative and is the least squares fit to the data. On its own, the nonnegative assumption has much experimental and theoretical evidence to support it and no contrary evidence.  In contrast, the combination of the nonnegative and least squares assumptions has little experimental evidence to support it and, as has been demonstrated in this paper, can yield false positive detections.

An argument that is often put forward in support of claims of the reliability of current practices is that the reduction in the reported myelin signal correlated with advanced pathology in postmortem formalin-fix brain (8, 21). Strong evidence has been presented for the correlation between the reduction in the myelin signal as determined with NNLS and increased pathology. However, as is well known, the increased pathology also has a major impact on the main water peak. As demonstrated in this paper, when reconstructing with NNLS, the shape of the main water peak can heavily influence the measured myelin signal. Crucially, the reconstructed

NNLS spectrum can give no indication that influence has taken place. Thus, the reported correlation between reduced myelin signal and advanced pathology could actually be detecting the well known correlation between advanced pathology and the main water peak.

**B. Testing the null hypothesis explicitly**

The explicit tests of the null hypothesis for the myelin detection problem clearly demonstrated the simplicity with which the test can be applied. That no prior information was required to implement the test enhances the confidence in any detection. The ability to include prior information, such as the nonnegative relaxation spectrum constraint, demonstrated the flexibility of the explicit test and the usefulness of the prior information, assuming it is valid.

The 500 SNR for a statistically significant detection of signal below 40ms, assuming the nonnegative constraint holds, places a lower limit on the SNR required for reliable measurement of the myelin signal. Measuring changes in the myelin signal will probably require more SNR. The 25 fold increase in the SNR required when the nonnegative relaxation spectrum constraint is removed clearly demonstrates the importance of the nonnegative constraint.

Explicit testing of the in vivo T2 decay yields similar results. The SNR of 650 allowed a detection of 4σ at 35.2ms.

**C. Testing the null hypothesis implicitly using reconstruction matrices**

Figures 5 and 6 show the biexponential T2 decays reconstructed with the reconstruction matrices for SNR of 1000 and 100 respectively. Figure 5 shows that for SNR of 1000 the myelin signal can be measured reliably if a nonnegative relaxation spectrum is assumed. However, Fig. 6 shows that for SNR of 100 the noise in the reconstructed spectrum is too large to reliably measure the myelin signal. Thus the results for the implicit implementation of the null hypothesis test bound those of the explicit results which yielded an SNR of 500 assuming the nonnegative assumption. The results may be slightly different as the explicit test uses a sharp cutoff at 40ms where the implicit test uses the weighted regions of the original spectrum provided by the resolution functions.

Figures 7 and 8, which are reconstructions of the broad peak spectrum, demonstrate that when there is no myelin signal in the original spectrum, the reconstruction matrices still yield reliable results, in contrast to the NNLS algorithm.

The reconstructed spectrum of the 48 echo *in vivo* T2 decay, which was reconstructed using data conserving reconstruction matrices, is presented in Cover (10). The reconstructed spectrum shows a signal detection below the 40ms time constant at about 5σ. This agrees well with the 4σ detection found when the explicit version of the null hypothesis test was applied to the 32 echo *in vivo* T2 decay. This also agrees with the 48 echo of the explicit version of the null hypothesis, but the offset correction implemented to handle the magnitude noise, leaves the results less definitive. Thus, for the *in vivo* T2 decay, the explicit and implicit version of the null hypothesis test are also in agreement.

A key property of the multiexponential reconstruction matrices is how well they handle the case where the data is consistent with both the existence and non existence of a particular signal. For example, as shown in Figure 2, some decays can be both consistent with the existence and non existence of the myelin signal. While reliable prior information maybe able to rule out one of these possibilities, it is important for an interpreter to understand whether the data on its own is consistent only with the existence or non existence of a particular signal. Spectra reconstruction with the reconstruction matrices makes this point clear to an interpreter.

The use of the nonnegative assumption only during the interpretation, and not during reconstruction, has the added advantage in the detection of artifacts. Since the reconstructed spectra should be nonnegative, any negative features indicate a deviation from the nonnegative assumption. Only artifacts, such as even-echo rephasing, should yields negative features. Thus, any negative features, other than noise, in a spectrum reconstructed with the data conserving reconstruction matrices are likely an indication of an artifact.

**D. Comparison with the FFT**

The data conserving multiexponential reconstruction matrices were claimed to fill the roll in multiexponential reconstruction that the FFT fills in frequency reconstruction. While detailed mathematical arguments have been presented previously justifying this claim (10) the results of this paper have re-enforced this claim by demonstrating the similarities in practice.

The only information used by both the FFT and the multiexponential reconstruction matrices is the information in the data. Examination of both algorithms shows they only incorporate the values and sampling information in the data. Also, they both allow trading off between resolution and SNR. In contrast, the most commonly used version of NNLS (version 2 in this paper) assumes there is no signal below 15ms. Removing this assumption, as was done in the first version of the regularized NNLS, yielded a myelin signal that went from nearly zero to half the expected value. The third version, which included the prior information that the relaxation spectrum had 4 discrete peaks, yielded a highly reproducible myelin signal. Thus, the prior information plays a major role in NNLS indicating it does not have the resistance to bad prior information assumptions shared by both the FFT and the reconstruction matrices. Still, the FFT and the data conserving reconstruction matrices both allow prior information to be invoked when interpreting the reconstructed spectra.

**E. Spectra reproducible with noise**

While the main purpose of this paper was to demonstrate the usefulness of extending the statistical test of rejecting the null hypothesis to relaxation spectrum space and its implicit use with the data conserving multiexponential reconstruction matrices, the results also raised questions about the reliability of measurements of the myelin signal in the literature.  As mentioned above, it is common practice currently to measure myelin signal in vivo using T2 decays with SNR close to 100 in combination with the NNLS reconstruction algorithm. However, as demonstrated in this paper, current practices can lead to highly reproducible false positive myelin signal detection.

**F. Summary of discussion**

Extending the concept of rejecting the null hypothesis to relaxation spectrum space has been shown to be a simple and robust method for detecting relaxation signals below 40ms in multiexponential decays. The avoidance of prior information in the statistical test allows errors due to unreliable prior information to be avoided. Using the multiexponential reconstruction matrices, which were data conserving, to implicitly test the null hypothesis made testing simple and intuitive to implement. The explicit and implicit implementation of the null hypothesis test where shown to yield similar results for both simulated decays and a decay measured *in vivo*. The results suggest that the multiexponential reconstruction matrices should play the same roll in reconstructing multiexponential spectra as the fast Fourier transform plays in reconstructing frequency spectra.

The explicit implementation of the null hypothesis in relaxation space also raises questions about the reliability of myelin signal measurements reported in the literature using current practices, where SNR not much above 100 for 32 echo with 10ms spacing is common. It was also demonstrated that T2 decays with similar sampling require SNR approaching 1000 for reliable detection and measurement of the myelin signal. While the susceptibility of NNLS to false positive myelin signal detection was demonstrated, other multiexponential reconstruction algorithms should be examined for similar susceptibilities. In addition, consideration should be given to reanalyzing the results of previous relaxation studies that original reconstructed T2 decays with the NNLS reconstruction algorithm with a technique not susceptible to false positive signal detections

**ACKNOWLEDGEMENTS**

Thanks to Hugo Vrenken, Jan C de Munck, Wolter L de Graaf and Bob W van Dijk for their helpful comments. The bulk of this research was carried out at the University of British Columbia. This research was supported primarily by D.W. Paty, D.K.B. Li, and the MS/MRI Research Group of the University of British Columbia. Additional support was received from the Netherlands' Virtual Laboratory for e-Science VL-e Project and the VU University Medical Center in Amsterdam.

**Table 1**
Confidence level of explicit rejection of the null hypothesis with no prior information for myelin signal detection. A confidence level exceeding 2σ is generally considered statistically significant.

| SNR | $\chi^2$ | Confidence level (σ) |
|---|---|---|
| 6,310 | 30± 8 | -0.3 |
| 7,943 | 35± 9 | 0.4 |
| 10,000 | 39±10 | 0.9 |
| 12,580 | 48±13 | 2.0 |
| 15,849 | 64±14 | 4.0 |
| 19,953 | 88±24 | 6.9 |
| 25,119 | 125±27 | 11.6 |
| 31,623 | 182±41 | 18.7 |

**Table 2**
Confidence level of explicit rejection of the null hypothesis for myelin signal detection including the prior information that the relaxation spectrum is nonnegative.

| SNR | $\chi^2$ | Confidence level (σ) |
|---|---|---|
| 200 | 34±10 | 0.2 |
| 251 | 34± 9 | 0.3 |
| 316 | 37±10 | 0.6 |
| 398 | 46±10 | 1.7 |
| 501 | 54±14 | 2.8 |
| 631 | 69±16 | 4.7 |
| 794 | 91±18 | 7.4 |
| 1,000 | 133±19 | 12.6 |

**Table 3**
Value of $\chi^2$ fit to a T2 decay of human white matter measured in vivo for relaxation spectrum with the spectrum forced to zero below the time constant in the first column. The $\chi^2$ values are for the first 32 and the full 48 points measured in the T2 decay.

| Time Constant (ms) | $\chi^2_{N=32}$ | $\chi^2_{N=48}$ |
|---|---|---|
| 1.0 | 28.8 | 192.0 |
| 10.0 | 29.7 | 193.5 |
| 11.2 | 29.7 | 193.8 |
| 12.6 | 29.9 | 194.1 |
| 14.1 | 30.7 | 194.6 |
| 15.8 | 32.2 | 195.4 |
| 17.8 | 34.3 | 196.4 |
| 20.0 | 37.2 | 197.8 |
| 22.4 | 41.0 | 200.0 |
| 25.1 | 45.5 | 204.4 |
| 28.2 | 50.9 | 211.4 |
| 31.6 | 58.1 | 221.3 |
| 35.5 | 67.2 | 233.5 |
| 39.8 | 78.0 | 247.9 |
| 44.7 | 91.2 | 268.5 |
| 50.1 | 107.2 | 295.1 |
| 56.2 | 137.9 | 337.1 |
| 63.1 | 202.9 | 405.8 |
| 70.8 | 338.7 | 532.2 |
| 79.4 | 721.0 | 884.2 |
| 89.1 | 3062.9 | 3255.4 |
| 100.0 | 10570.1 | 10880.4 |

**Figures**

**Figure 1.** The original spectrum referred to as the broad peak spectrum. It was used to generate some of the simulated decays for reconstruction.

**Figure 2.** Three versions of the NNLS reconstruction of the relaxation spectra from a 32 echo simulated decay generated from the broad peak decays with SNR of 100. The false positive myelin signal is highly reproducible when there are only 4 discrete time constants (Fig 2(c)).

**Figure 3.** Point spread functions for the seven multiexponential reconstruction matrices used in this paper. As expected, the resolution increases with the noise gain (G) of the matrix.

**Figure 4.** Resolution functions of the same seven reconstruction matrices as displayed in Fig3.

**Figure 5.** Reconstruction of the simulated biexponential decay with SNR of 1000. The myelin signal is clearly above the noise at 20ms for G=3.16.

**Figure 6.** Reconstruction of the simulated biexponential decay with SNR of 100. The myelin signal cannot be seen because of the noise.

**Figure 7.** Reconstruction of the simulated decay from the broad peak original spectrum with SNR of 1000. Any signal at 20ms is clearly much smaller than that expected of myelin.

**Figure 8.** Reconstruction of the simulated decay from the broad peak original spectrum with SNR of 100. The reconstruction is clearly too noisy to determine if a myelin signal with the expected amplitude is present or not.

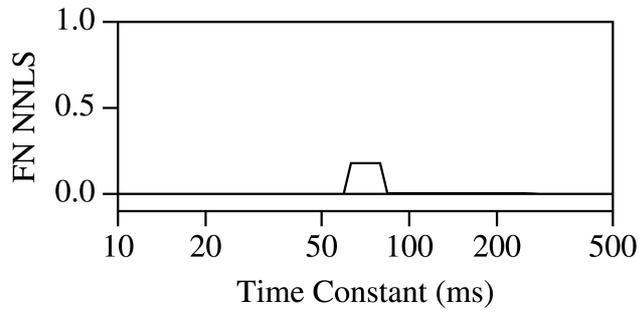

**Figure 1**

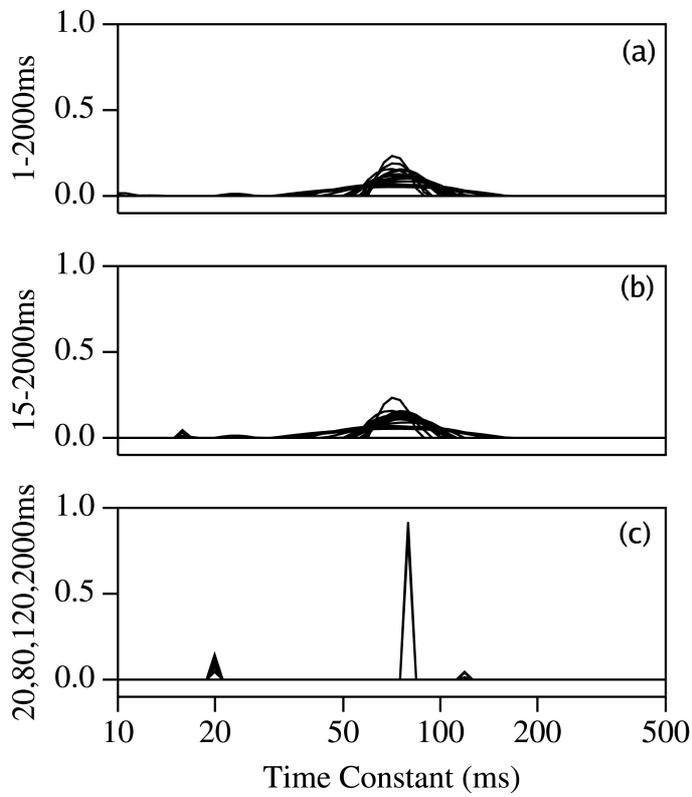

**Figure 2**

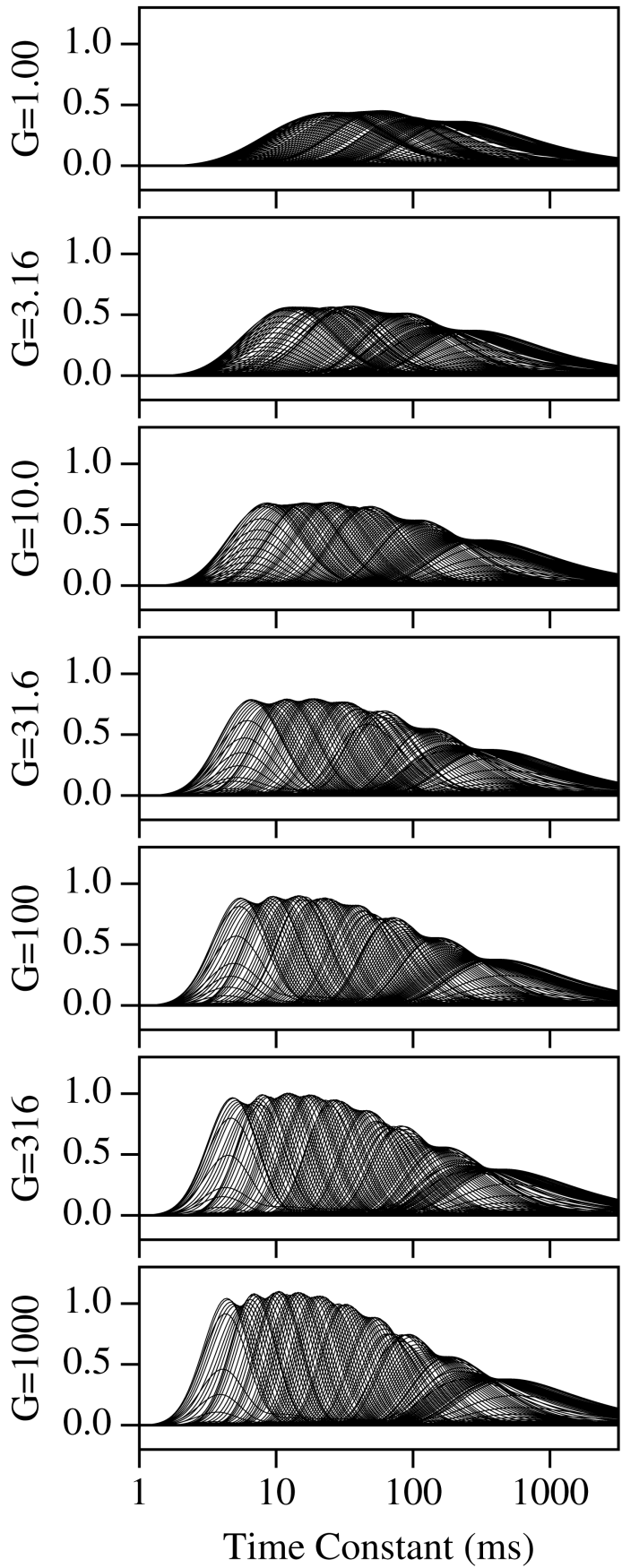

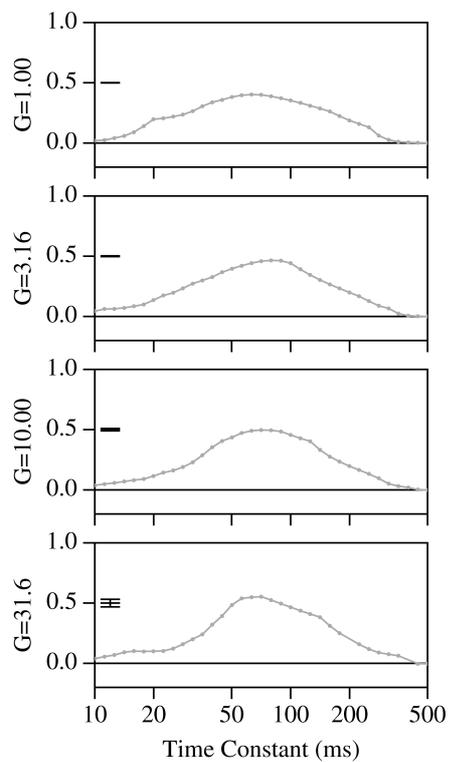

**Figure 5**

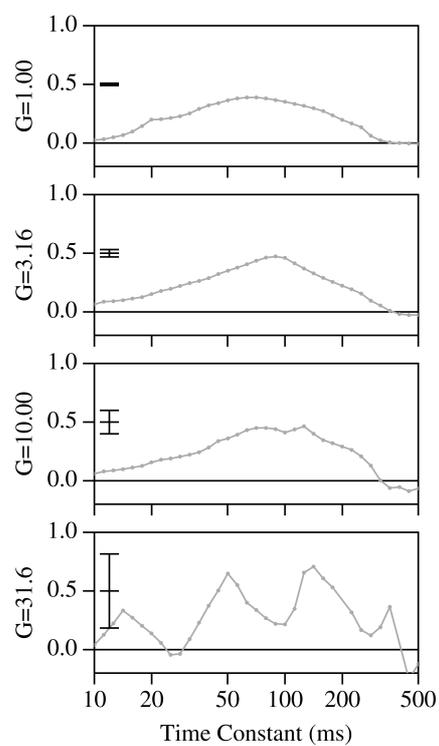

**Figure 6**

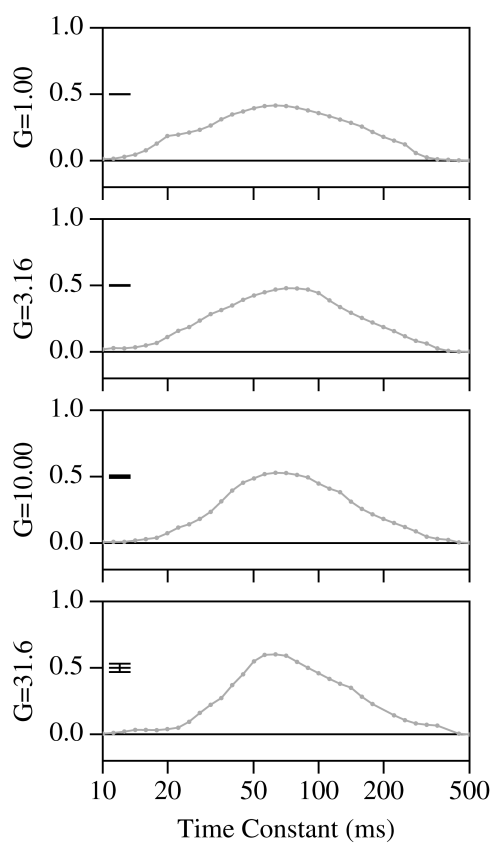

**Figure 7**

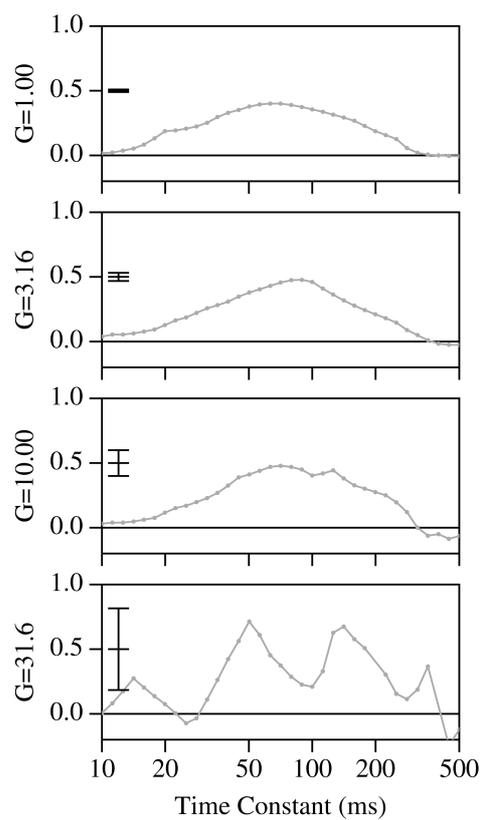

**Figure 8**